\documentclass[a4paper]{jpconf}
\usepackage{graphicx}

\def\apj{ApJ}%
\def\apjl{ApJ}%
\def\aap{A\&A}%
\begin{document}
\title{Surface detonation in type Ia supernova explosions?}

\author{F.~K.~R{\"o}pke$^{1,2}$ and S.~E.~Woosley$^1$}

\address{$^1$University of California at Santa Cruz, 1156 High Street,
  Santa Cruz, CA 95060, U.S.A}
\address{  $^2$Max-Planck-Isntitut f{\"u}r Astrophysik, Karl-Schwarzschild-Str.~1,
  85741 Garching, Germany}

\ead{roepke@ucolick.org}

\begin{abstract}
We explore the evolution of thermonuclear supernova explosions when
the progenitor white dwarf star ignites asymmetrically
off-center. Several numerical simulations are carried out in two and
three dimensions to test the consequences of different initial flame
configurations such as 
spherical bubbles displaced from the center, more complex deformed
configurations, and teardrop-shaped ignitions. The burning bubbles
float towards the surface while releasing energy due to the nuclear
reactions. If
the energy release is too small to gravitationally unbind the star,
the ash sweeps around it, once the burning bubble approaches the
surface. Collisions in the fuel on the opposite side increase its
temperature and density and may -- in some
cases -- initiate a detonation wave which will then propagate inward
burning the core of the star and leading to a strong explosion. However, for
initial setups in two dimensions that seem realistic from pre-ignition
evolution, as well as for all three-dimensional simulations the
collimation of the surface material is found to be too weak
to trigger a detonation. 

\end{abstract}

\section{Introduction}
Type Ia supernovae are generally
associated with the thermonuclear 
explosion of earth-sized white dwarf (WD) stars -- the final stages of
the evolution of small and intermediate-mass stars. These dense
objects consist of carbon and oxygen and
their pressure is dominated by 
degenerate electrons. In a binary system, the WD can gain mass
from its companion by accretion.
However, there exists a fundamental limit, the Chandrasekhar mass,
beyond which the star is unstable against
gravitational collapse. Before reaching this stage, the density at the
center of the WD reaches values at
which nuclear reactions from carbon to heavier elements ignite. This
establishes a stage of convective
carbon burning which is finally terminated when a thermonuclear runaway forms a
flame. The thermonuclear flame propagates outward and may give rise to
an explosion of the star \cite{reinecke2002d,roepke2005b}. The process resembles
premixed chemical combustion where two solutions -- supersonic
detonation and subsonic
deflagration -- are admissible.
From observational constraints it is clear that the flame starts out
as a subsonic deflagration. The ashes are buoyant and shear instabilities
generate strong turbulence. The interaction of the flame with
turbulence is a key feature in the model
and accelerates the flame. Consequently, one of the outstanding
problems is to determine, whether this turbulent flame leads to
explosions energetic enough to be consistent with observations, or
whether a later transition to a supersonic detonation is both possible
and necessary. In the freely expanding medium of SN Ia explosions,
such transitions are difficult to achieve.

The exact way of flame ignition
is not well determined yet, but previous numerical studies
\cite{woosley2004a, kuhlen2006a} suggest
that the ignition process  
may proceed on only one side of the star's center in a few
patches. While it has been shown that ignitions in
multiple spots distributed around the center of the WD lead to strong
explosions \cite{roepke2006a}, one-sided ignitions are expected to release less energy.
The effects of the latter scenario on the 
explosion process are investigated in the present study.
It has been speculated that collisions between gravitationally bound laterally expanding
regions of fuel and ash could, in principle, give rise to a
``gravitationally confined detonation'' \cite{plewa2004a}. This possibility
is studied in detail.

\section{Numerical model and computation}

The code used for simulating the thermonuclear supernova explosions on
scales of the WD star has been described in detail previously \cite{reinecke1999a,reinecke2002b,
roepke2005c}. Since it is not possible to resolve all relevant spatial
scales of the problem, an approach similar to Large Eddy Simulations
was chosen.

The modeling of the hydrodynamics  is based on the \textsc{Prometheus} \cite{fryxell1989a}
implementation of the piecewise parabolic method \cite{colella1984a}. Our
flame model  
consists of three components. Flame propagation is followed
via a level-set approach. It associates the flame with the zero level set
of a signed distance function $G$ which is evolved in a suitable way
\cite{reinecke1999a} to account for the propagation. This flame
prescription, however, does not resolve
the scales that determine the propagation velocity. In the flamelet
regime of turbulent combustion, which applies to the situation in
thermonuclear supernovae (possibly with the exception of the very
late stages, e.g.\ \cite{roepke2005a}) the flame velocity is proportional to the turbulent
velocity fluctuations. These are determined from a turbulent
subgrid-scale model \cite{schmidt2006a} featuring localized closures
to take into account the non-isotropic nature of the problem.  The
energy release due to nuclear
burning is handled in a simplified way. Only five species are
considered: carbon and oxygen of the initial WD material, $^{24}$Mg
representing intermediate mass elements, alpha particles, and $^{56}$Ni
as a representative of iron group nuclei. At fuel densities higher
than $5.25 \times 10^7 \, \mathrm{g} \, \mathrm{cm}^{-3}$, the
material crossed by the flame is burned to
nuclear statistical equilibrium, modeled as a mixture of alpha
particles and $^{56}$Ni. Below, this density burning is incomplete and
produces intermediate mass elements. The appropriate amounts of energy
are released in the ashes. At densities below $10^7 \,
\mathrm{g} \, \mathrm{cm}^{-3}$ the reactions cease.

The code is efficiently parallelized on the basis of domain
decomposition in an MPI implementation. While two-dimensional
simulations can be run easily on a multi-processor desktop computer,
three-dimensional simulations and require a supercomputer.
The code has been used on several machines and proved good scaling
behavior. The three-dimensional simulations described here were carried
out on a $[512]^3$
cells computational grid required about 20,000 CPU hours each on the
NCCS Jaguar Cray XT3 supercomputer. Simulations of this size were
performed on 256 processors.

\section{Simulations}

\begin{figure}[t]
\centerline{\includegraphics[width=0.8\textwidth]{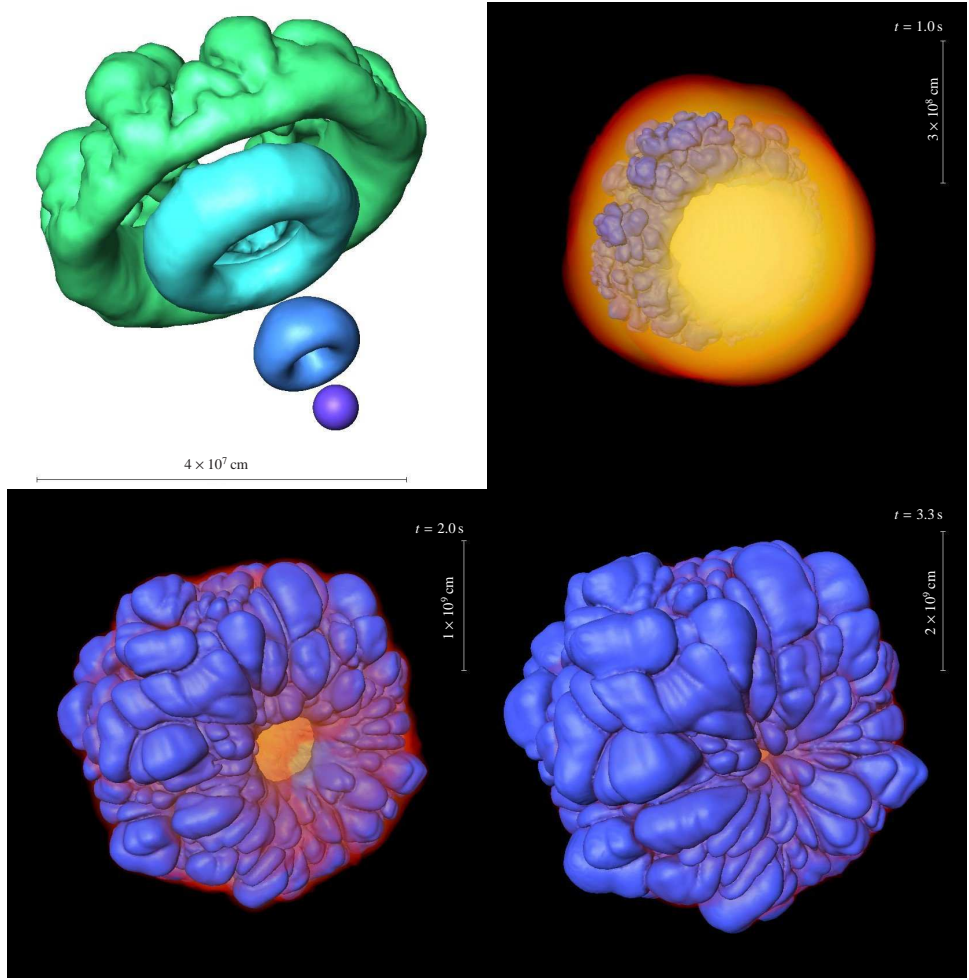}}
\caption{\label{fig:evo} Evolution of an explosion simulation with the flame ignition
 in a single bubble of radius $25 \, \mathrm{km}$ displaced $100 \,
 \mathrm{km}$ from the center of the WD. Top left panel: initial
 evolution of the flame front (blue to green isosurfaces correspond to
 $t = [0, 0.25, 0.35, 0.45] \, \mathrm{s}$). Other panels: later evolution with the
 logarithm of the density volume rendered and $G=0$ as blue isosurface
 indicating the flame front or, later, the approximate boundary
 between burnt and unburnt material.}
\end{figure}

In our model, the flame is artificially ignited. Therefore the initial
flame configuration is an undetermined parameter. Since it is not
constrained tightly by complementary studies, several numerical
experiments with asymmetric ignitions have been performed.
We considered initial flame configurations in form of a single
spherical bubble placed at different radii off-center of the WD, 
more complex  bubble-substructures and perturbations from perfect
sphericity, and (on average) teardrop-shaped initial flames extending
down to the center of the star or even overshooting it.

In Fig.~\ref{fig:evo}, the deflagration flame was ignited in a single spherical
bubble 200 km off-center. This bubble is
filled with hot ashes of lower density than the surrounding fuel and therefore  is
buoyant. As it burns and floats,
the flame shape evolves from a sphere to a torus
in the first tenths of a second (upper left panel of
Fig.~\ref{fig:evo}). Continuing towards the surface, the torus
becomes distorted by growing features that eventually connect.

The 
energy released during the flame's transit from the center to the surface
is too small to explode 
the WD, which remains gravitationally bound.
Once the bubble of ash breaks out, it starts to sweep around the core,
colliding on the diametrically opposite side of the star. In the
collision, the temperature and the density of unburnt material are
increased.

The question is whether this increase is sufficient to initiate
a detonation. A detonation wave would propagate 
inwards burning the core of the star and leading to an 
explosion. Clearly, the strength of the collision depends upon the
initial parameters mentioned above, all of which alter the energy release in the
nuclear burning and hence the expansion of the WD prior the the
surface material collision. These parameters were tested in about 20
two-dimensional simulations. Since such simulations impose an artificial
cylindrical symmetry and, moreover, turbulence behaves differently in
2D and 3D simulations, neither the nuclear energy release nor the collision
strength are expected to be realistic. Therefore, the 2D simulations
merely provide a way of exploring the dependencies. Five
three-dimensional simulations were carried out in addition to assess
what might realistically happen in a supernova.

\section{Results and conclusions}

\begin{figure}[t]
\includegraphics[width=0.7\textwidth]{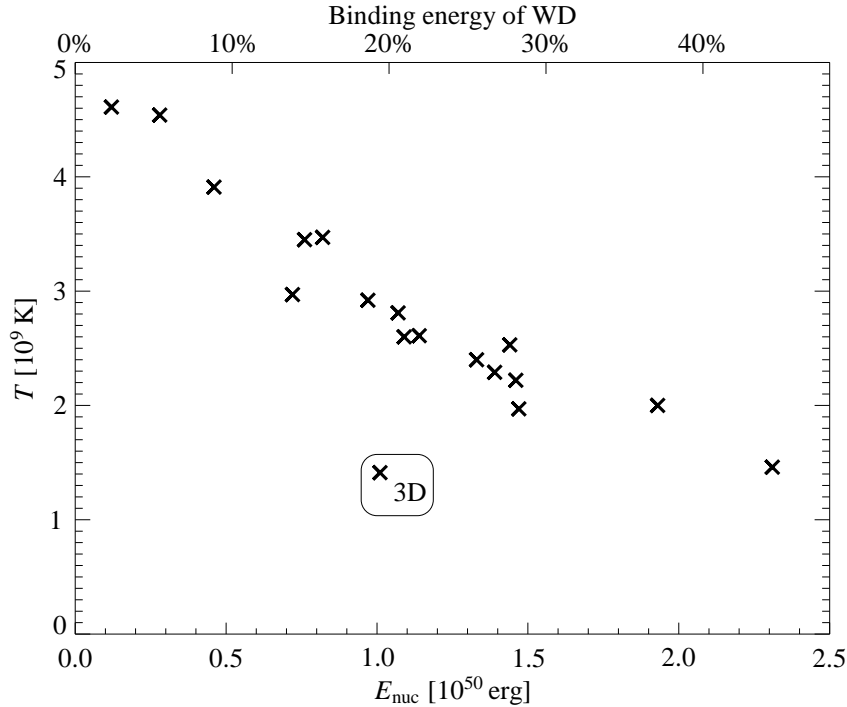}
\hspace{0.01 \textwidth}
\begin{minipage}[b]{0.299\textwidth}
\caption{\label{fig:et1} Maximum fuel temperature reached in the region of
  collision of the material sweeping around the WD as a function of
  the nuclear energy release prior to reaching that temperature. The
  3D data point is marked, all other points correspond to 2D
  simulations.\vspace*{6ex}}
\end{minipage}
\end{figure}

Initiation of a detonation wave in the collision region depends on
the temperature reached, the density, and the spatial extent of the
collision region.
For our 2D simulations ignited in a spherical bubble, an inverse
correlation exists between nuclear energy
release prior to collision and the maximum temperature reached (cf.\
Fig.~\ref{fig:et1}). This correlation arises naturally from the expansion of the
star due to the energy release. The more material is burnt in the
flame on its way to the surface, the more the star expands and the
weaker is the collision.

The energy release depends on the displacement of the initial flame
from the center. Initiating the flame at larger radii leads to less
burning and thus to stronger collisions of the surface material.
The collision strength is very sensitive to perturbations of the
initial flame from perfect sphericity, either directly by imposing
more complex initial configurations (teardrop-like shapes, multiple
initial flame kernels) or by different numerical resolution and
resulting discretization errors.

\begin{figure}[t]
\includegraphics[width=0.7\textwidth]{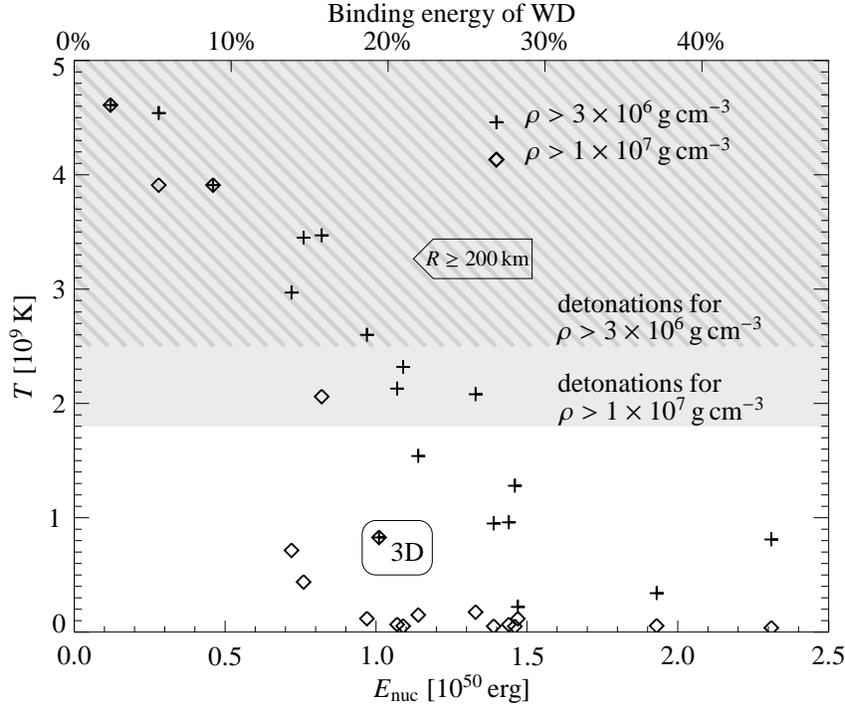}
\hspace{0.01 \textwidth}
\begin{minipage}[b]{0.299\textwidth}
\caption{\label{fig:et2} Maximum temperature reached in the collision
  region for fuel above the given density thresholds as a function of
  the nuclear energy release prior to reaching the maximum collision
  temperature. The shaded and dashed regions correspond to conditions
  where an initiation of a detonation is possible for temperatures
  reached in fuel of densities $\rho > 1 \times 10^7 \, \mathrm{g} \,
  \mathrm{cm}^{-3}$ and $\rho > 3 \times 10^6 \, \mathrm{g} \,
  \mathrm{cm}^{-3}$, respectively.\vspace*{6ex}}
\end{minipage}
\end{figure}

To trigger a detonation, a collision temperature over
$2 \times 10^9 \, \mathrm{K}$ is necessary. However, this threshold
temperature needs to be 
reached in fuel with density higher than $\sim 3\ldots10\times 10^6 \,
\mathrm{g} \, \mathrm{cm}^{-3}$. Therefore, the
maximum temperature in the collision region at a density exceeding
these values was measured. The results are plotted in Fig.~\ref{fig:et2}. 
From this measurement, we find that a detonation initiation is only
possible in those 2D models that start with a spherical initial flame placed
at unrealistically large radii ($R \ge 200\, \mathrm{km}$) off-center
of the WD star (cf.\ Fig.~\ref{fig:et2}).

3D simulations are cooler in the collimation region for the
same nuclear energy release. However, except for one model (included
in Figs.~\ref{fig:et1} and \ref{fig:et2}) the 3D simulations release
more energy in the nuclear burning than the 2D simulations.
In the model shown in Fig.~\ref{fig:evo}, the energy release amounted
to $\sim$$2.8 \times 10^{50} \, \mathrm{erg}$ and the temperature in the collision
region barely exceeded $10^9 \, \mathrm{K}$. For the one-sided
teardrop-shaped initial flame the energy release was even greater and
one singe-bubble ignition and the two-sided teardrop-ignition unbound
the star. All 3D simulations clearly failed to trigger a detonation.

We conclude, that although a detonation due to the colliding surface
material is in principle possible for specific (and artificial) ignition
configurations, it is unlikely to happen in nature.
For models that remain gravitationally bound, failures to initiate a
detonation  will lead to pulsations of the WD star. This may be a
second chance for triggering a detonation ("pulsational delayed
detonation scenario", \cite{arnett1994b}) and will be addressed in a follow-up study.

\ack{This research used resources of the National Center for
  Computational Sciences at Oak Ridge National Laboratory, which is
  supported by the Office of Science of the U.S. Department of Energy
  under Contract No. DE-AC05-00OR22725.}


\end{document}